\documentclass[aps,twocolumn,superscriptaddress,showpacs,showkeys,pra,amssymb, amsmath]{revtex4-1}
\usepackage{amsmath,latexsym,amssymb,amscd}
\usepackage{graphicx}
\usepackage{amsmath}
\usepackage{latexsym} 
\usepackage{amssymb} 

\newtheorem{theorem}{Theorem}
\newtheorem{thm}[theorem]{Theorem}

\newtheorem{lem}[theorem]{Lemma}

\begin{document}
\title{Nonparametric Regression Quantum Neural Networks}
\date{\bf \today}

\author{Do Ngoc Diep}
\address{TIMAS, Thang Long University, Nghiem Xuan Yem Road, Hoang Mai district, Hanoi, Vietnam}
\address{Institute of Mathematics, Vietnam National Academy of Science and Technology, 18 Hoang Quoc Viet road, Cau Giay district, 10307 Hanoi, Vietnam}

\email{diepdn@thanglong.edu.vn}

\author{Koji Nagata}
\affiliation{Department of Physics, Korea Advanced Institute of Science and Technology, Daejeon 34141, Korea}

\author{Tadao Nakamura}
\affiliation{Department of Information and Computer Science,  
Keio University, 3-14-1 Hiyoshi, Kohoku-ku,Yokohama 223-8522, Japan}

\date{\textbf{\today}}

\begin{abstract}
\textbf{Abstract}\\  In two pervious papers \cite{dndiep3}, \cite{dndiep4}, the first author constructed the least square quantum neural networks (LS-QNN), and ploynomial interpolation quantum neural networks ( PI-QNN),  parametrico-stattistical QNN like: leanr regrassion quantum neural networks (LR-QNN), polynomial regression quantum neural networks (PR-QNN), chi-squared quantum neural netowrks ($\chi^2$-QNN). We observed that the method works also in the cases by using nonparametric statistics.
In this paper we analyze and implement the nonparametric tests on QNN such as: linear nonparametric regression quantum neural networks (LNR-QNN), polynomial nonparametric regression quantum neural networks (PNR-QNN). The implementation is constructed through the Gauss-Jordan Elimination quantum neural networks (GJE-QNN).The training rule is to use the high probability confidence regions or intervals.
\end{abstract}
\maketitle
\section{Introduction}

The classical machine learning (ML) theory, see e.g.  \cite{hinton} was created in 1950 as some systems of network computing, then  9 years later  in 1959 Arthur Samuel gave a definition as some unpredictedly programmed computing.  The functions (as input-output boxes) are deduced from a set of training data. The classical ML is characterized with four ingredients: - \textit{supervised learning},  - \textit{unsupervised learning}, - \textit{training} and - \textit{reinforcement learning}. As a rule, the classical MLs are working with \textit{big data}.

The \textit{quantum Machine Learning} (QML) \cite{hinton}-\cite{ezhovventura} are characterized by using quantum computing into the theory. One uses the ordinary interpretation of qubits, 1-qubit quantum gates, such as the Pauli matrices, etc. 
$\mathbf 1 = -\fbox{Id}- \sim \begin{pmatrix} 1 & 0\\ 0 & 1 \end{pmatrix},$ 
$X= -\fbox{X}-\sim \begin{pmatrix} 0 & 1 \\ 1 & 0\end{pmatrix},$ 
$Y = -\fbox{Y}-\sim \begin{pmatrix} 0 & -i \\ i & 0 \end{pmatrix},$ $Z = -\fbox{Z}-\sim\begin{pmatrix} 1 & 0\\ 0 & -1 \end{pmatrix},$ then the 2-qubit gates like $\mathrm{XOR} = -\fbox{XOR}- \sim \begin{pmatrix} 1 & 0 &  0 & 0\\
0 & 1 & 0 & 0\\ 0 & 0 & 0 &1\\ 0 & 0 & 1 & 0\end{pmatrix},$  $\mbox{SWAP} = -\fbox{SWAP}- \sim  \begin{pmatrix} 1 & 0 & 0 & 0\\ 0 & 0 & 1 &0\\ 0 & 1 & 0 & 0\\ 0 & 0 & 0 & 1 \end{pmatrix},$ and finally,  Measurements $\mathrm{M}=-\fbox{M}-. $

Quantum algorithms are used to solve the ML problems with quantum computing. The most important ingredients in QML are:  - choices of  \textit{training sets}, i.e. finite sets of given vectors in order to then find some value corresponding to another input, - \textit{pattern completion}, which means adding missing informations to incomplete inputs, and  - \textit{associative memory}, i.e. retrieving  stored memory vectors upon an input.

Following the Deutsch's model, a \textit{quantum neural network} $\mathrm{QNN}(s,d)$ is a set of all quantum circuits of size $s$ and depth $d$ with thresholds bounded by $w$. Quantum gates are interconnected by wires, preserve the sources and sink gates measured the qubits and removed the entanglements with the maining qubits. Examples of QNNs \cite{guptazia} are the implementation of NAND gate, dissipative gates $D(m,\delta)$ and sink gates $X$.

A \textit{threshold circuit} is a boolean function $Th^{n,\Delta}: \mathbb Z_2^n \to \mathbb Z_2$ of $n$ integral variables $x_1,\dots,x_n$ such that $Th^{n,\Delta}(x_1,\dots,x_n) = 1$ if and only if $\sum x_iw_i \geq \Delta$. The class $TC(s(n),d(n))$ of threshold circuits of size $s(n)$ and depth $d(n)$, weighted by weight bound $w$ can be approximated by elementary functions.  

An \textit{equality (threshold) circuit} is a boolean function $Et^n_{w_1,\dots,w_n}: \mathbb Z_2^n \to \mathbb Z_2$ of $n$ integral variables $x_1,\dots,x_n$ such that $Et^n_{w_1,\dots,w_n}(x_1,\dots,x_n) = 0$ if and only if $\sum x_i w_i=0$. The class $EC(s(n),d(n))$
 of equality threshold circuits of size $s(n)$ and depth $d(n)$, weighted by weight bound $w$ can be approximated by elementary functions.  
 
 It was proven that $TC(s(n),d(n)) \subseteq EC(O(s^2(n),2d(n))$ of weight bound $O(s(n))$ and $TC(s(n),d(n)) \subseteq EC(O(s^2(n),d(n)+1)$ of weight bound $O(s^2(n))$. And finally, $EC(s(n),d(n)) \subseteq QNN(O)d(n).\log s(n)), 2d(n))$ of precision $O(\log w + d(n)\log s(n))$.
 (Theorem 4.6 from \cite{guptazia}).
  
In two pervious papers \cite{dndiep3}-\cite{dndiep4}, the first author constructed the least square method quantum neural networks (LS-QNN), the polynomial interpolation quantum neural network (PI-QNN) , the parametrico-stattistical QNNs like:  linear regrassion quantum neural networks (LR-QNN), the polynomial regression quantum neural networks (PR-QNN),and the chi-squared test quantum neural networks ($\chi^2$-QNN).
  
In this paper we analyze and implement the nonparametric tests on QNN. In place of high probability region of  standardized test statistics, we use the  band $\mathbb P(a(x) \leq r(x) \leq b(x))=1-\alpha$ to implement the training rules.

The structure of the paper is as follows. In the next Section II we describe the main topics of nonparametric statistics. We then construct the linear nonparametric regression (LNR-QNN), namely U test QNN, .... in the next Section III and then the polynomial nonparametric regression quantum neural networks (PNR-QNN) in the Section IV. The paper will be finished with conclusion in Section V and Acknowledgments in Section VI.

\section{Nonparametric statistics}
The main topics \cite{wasserman} of nonparametric statistics  are as follows.\\
1. One starts with the \textit{problem of estimating} the distribution function for sample  $X_1, \dots, X_n \sim F$, estimate the commulative distribution function (cdf) $F (x) = \mathbb P(X\leq x)$
\\
2. One tries to find out some \textit{estimating functionals}: given a sample  $X_1, ... , X_n \sim F$, estimate a functional T(F) such as the mean $T(F) = \int xdF(x)$
and the  \textit{Density estimation} $f(x) = F'(x)$. \\
3. One then makes  a \textit{nonparametric regression or curve estimation}: given points $(x_1,y_1),...,(x_n,y_n)$ estimate the regression function $r(x) = \mathbb E(Y |X = x)$. \\
4. Finally, one estimates the \textit{unknown normal means}: given $Y_i \sim  \mathcal N(\theta_i,\sigma^2), i = 1,\dots,n$, estimate $\theta=(\theta_1,\dots,\theta_n)$. \\
Also some miscellaneous problems is such as measurement error, inverse problems and testing should be considered.

Let us consider the regression problem in nonparametric statistics.
Given $n$ points $(x_1,y_1), ... ,(x_n,y_n)$ associated to a probability distribution $F (x) = \mathbb P(X\leq x)$ of an independent random variable $X$ and a dependent variable $Y$, we try to find a regression function $r(x)= \mathbb E(Y |X = x)$. In the parametric regression, we know the type of function $r(x)$, namely linear or polynomial or some exponential type, etc... In nonparametric regression, we do not know the type of the distribution. 

The \textit{linear model of regression} is as follows. Let $y_i = r(x_i) + \varepsilon_i, i=1,\dots,n$, with assumption that the random variables $\varepsilon_i$ have  vanishing expected values $\mathbb E(\varepsilon_i) = 0, \forall i$ and the unknown constant variances $\mathbb V(\varepsilon_i) = \sigma^2$. Following the least square method, the solution to the problem is $r(x) = \beta^T.x= \sum_{j=1}^p \beta_jx_{j}$, with coefficients $\beta = (\beta_1,\dots,\beta_p)^T$. For $p$ sample $X_1,\dots,X_p$, denote by $$X= [X_1,\dots, X_p] = \left[ \begin{matrix} x_{11} & x_{12}& \dots & x_{1p}\\
x_{21} & x_{22} & \dots & x_{2p}\\
\dotfill & \dotfill & \dotfill & \dotfill\\ 
x_{n1} & x_{n2} & \dots & x_{np}
\end{matrix}\right]$$ the \textit{hat matrix}, the entries of the $i^{th}$ column of which are the coordinates of sample $X_i$. The linear regression equation becomes $$Y = X\beta + \varepsilon.$$ Let $\hat\beta = (\hat\beta_1,\dots, \hat\beta_p)^T$ be the column of estimators of the column $\beta$, that minimizes the residual sum of squares
$$RSS =(Y-X\beta)^T(Y-X\beta) \to\min.$$ The column-vector estimator  $\hat\beta$ is the solution to the dual equation
$$(X^TX)\hat\beta = X^TY. \eqno{(2.1)}$$ In general, following the Kronecker's condition, the equation (2.1) is consistent if and only if the last column $X^TY$ appeared in the system (2.1) augmented matrix $[(X^TX)\; X^TY]$ is  a non-pivot column. In that case we can find all the solution $\hat\beta$ by the Gauss-Jordan Elimination Procedure. If the regression points $(X_i,Y_i), i=1,\dots,p$ are in generic position then really $X^TX$ is invertible. In that case the solution $\hat\beta$ is unique and can be found in form $$\hat\beta = (X^TX)^{-1}X^TY.$$
In the general case, the matrix $X^TX$ could be degenerated, one uses the Moore-Penrose pseudo-inverse $(X^TX)^{-1}_{psi}$ to find the solutions in the form
$$\hat\beta = (X^TX)^{-1}_{psi}X^TY.$$
The regression function $r(x) = \sum_{j=1}^p \hat\beta_jx_j = x^T\hat\beta$ has also estimators $\hat r(x) = X\hat\beta$, which can be considered as solution to the equation
$$ x^T(X^TX)\hat\beta = x^TX^TY. \eqno{(2.2)}$$ 
Again in general one solves the system by the method of Gauss-Jordan Elmination with the Kronecker's condition for the last column of the system augmeted matrix not to be pivot.
In the case of generic points of regression, i.e. the matrix $X^TX$ is invertible, one introduced the symmetric and idempotent matrix $L= X(X^TX)^{-1}X^T$, the trace of which $p={\mathrm tr}(L)$ is the effective degree of freedom.
The unbiased estimator of $\sigma^2$ is $$\hat\sigma^2 = \frac{1}{n-p}\sum_{i=1}^n (y_i-r_n(x_i))^2 = \frac{||\hat\varepsilon||^2}{n-p}. \eqno{(2.3)}$$
The problem can be stated as follows. For a fixed level of confidence $\alpha$ find two function $a(x)$ and $b(x)$ such that they bound the high probability region 
$$\mathbb P(a(x) \leq r(x) \leq b(x))) = 1-\alpha. \eqno{(2.4)}$$ In the next section we will use this condition to provide the training rule in quantum neural networks.

Let $\ell_i(x) $ be the entries of the column  $\ell(x)$ defined from the equation (2.2). The variances of enstimator $\hat r_n(x)$ is 
$$\mathbb V(\hat r_n(x)) = \sigma^2 \sum_{i=1}^n \ell_i(x) = \sigma^2 ||\ell(x)||^2. $$ The band of high probability $1-\alpha$ is of the form $$I(x) = (a(x),b(x)) = $$ 
$$ (\hat r_n(x) -c\hat\sigma ||\ell(x)||, \hat r_n(x) + c\hat\sigma ||\ell(x)||)$$ 
for an appropriate constant $c$. 
As in the parametric case, in nonparametric statistics, one uses also the $F$-ratio $F_{p,n-p}$ to compare with the critical $F$-ratio $F_{\alpha; p,n-p}$. 
We refer to the following basic result of Wasserman \cite{wasserman}:
\textit{For any confidence level $\alpha$, the confidence interval $I(x) = (a(x),b(x)) = (\hat r_n(x) -c\hat\sigma ||\ell(x)||, 
\hat r_n(x) + c\hat\sigma ||\ell(x)||)$ satisfies the condition 
$$\mathbb P(a(x) \leq r(x) \leq b(x))) = 1-\alpha $$ for constant $c = \sqrt{pF_{\alpha; p,n-p}}$}

\section{Linear Nonparametric Regression Quantum Neural Networks}

Let us recall a result from \cite{dndiep1}, which plays a fundamental role in implementation of statistical test QNNs.
\begin{lem}
The Quantum Gauss-Jordan Elimination (QGJE) code can be implemented in QNN.
\end{lem}
Indeed, in \cite{dndiep2} we reminded the implementation of XOR or CNOT gates in QNN. 
The Grover Search is repeatedly implemented by Hadamard gates, function-evaluation query gates. After that,
the QGJE code in to use the Grover's Search for finding the pivot columns in a matrix.

 In QNN one starts with the elementary gates to provide the Grover's Search code and then to implement the QGJE code:
$$\CD \mbox{XOR, SWAP, H, X,Y,Z}, \mathbf 1 @>>> \mbox{Grover's Search}\\ @>>> \mbox{QGJE} \endCD$$

The least squared quantum neural networks LS-QNN can be implemented in quantum neural networks by using the Gauss-Jordan Elimination Code \cite{dndiep1}.

\textit{The training rule is provided if it is located in the confidential interval $I(x)$ above}. We therefore have the following result

\begin{thm}
The linear nonparametric regression quantum neural networks LNR-QNN can be implemented by using the Gauss-Jordan Elimination quantum neiral network Code GJE-QNN.
\end{thm}

Indeed, the XOR gates are implemented on QNN's as XOR-QNN. It was shown in \cite{dndiep3}, \cite{dndiep4} that the quantum Fourier transforms $\mathcal F_n$ are also implemented on quantum neural networks (QF-QNN). The composition $\mathcal F_n \circ Q_{f_0} \circ \mathcal F^\dagger_n$ of Fourier transforms $\mathcal F_n$ followed by the initial query $Q_{f_0}$ and the inverse quantum Fourier transform $\mathcal F^\dagger_n$, then followed by
the repeated compositions $\mathcal F_n \circ Q_{f} \circ \mathcal F^\dagger_n$  of Fourier transforms with query $Q_f$ and the inverse  Fourier transforms, repeated $d^{n/2}$ times, give the implementation of Grover's Search on QNN \cite{dndiep3}. The quantum GJE-QNN is effectively implemented then by using the Grover's Search quantum neural networks (GS-QNN) for finding the pivot columns \cite{dndiep3} of linear systems.  
\vskip 0.25cm

\textbf{Remark}. By the same way we can also implement the logistic local regression on quantum neural network.

\section{Polynomial Nonparametric  Regression Quantum Neural Networks}

The problem is again to find the regression function $r(x)$ by local polynomial on $n$ variables of degree $p$, see \cite{wasserman} for the case $n=1$:
$$P_x(u,a) = \sum_{|\alpha|=0}^p a_\alpha(u-x)^\alpha, \eqno{(4.1)}$$ where the multi-index notation $\alpha= (\alpha_1,\dots,\alpha_n)$, $(u-x)^\alpha = \Pi_{|\alpha|=0}^p (u_1-x_1)^{\alpha_1} \dots (u_n-x_n)^{\alpha_n}$.

\begin{thm}
Local kernel nonparametric regression quantum neural networks KNR-QNN can be implemented by using the Gauss-Jordan Elimination quantum neural network GJE-QNN.
\end{thm}

Indeed, let us consider the estimator $a_i\equiv \hat r(x_{(i)})$ as the estimator to the weighted sum of squares 
$$\sum_{|i|=0}^p w_i(x) (y_i-a_i)^2 \to \min. \eqno{(4.2)}$$
One uses $$\hat r(x) = \frac{\sum_{i=0}^p w_i(x)y_i}{\sum_{i=0}^n w_i(x)}$$
$$\hat r(u) \approx P_x(u,\hat a)$$ is the regression polynomial.
Then $\hat r_p(x) = P_p(x-x;\hat a) = \hat a_0(x)$. Let  us use the mutiindex notation to consider the hat matrix $X_x$ at any $x$.
$$X_x= \left[   
\begin{matrix} 1 & (x_{(0)}-x)^{(1,\dots,0)} & \dots & \frac{(x_{(0)}-x)^{(0,\dots,p)}}{p!} \\ 
1 & (x_{(1)}-x)^{(1,\dots,0)} & \dots & \frac{(x_{(1)}-x)^{(0,\dots,p)}}{p!} \\ 
\dotfill &\dotfill & \dotfill &\dotfill \\
1 & (x_{p}-x)^{(1,\dots,p)} & \dots & \frac{(x_{p}-x)^{(0,\dots,p)}}{p!} 
\end{matrix}
\right], \eqno{(4.3)}$$
and let $$W_x = \mathrm{diag}(w_0(x), \dots, w_p(x))$$
then for $Y = [y_0,\dots, y_p]^T$ we can always suppose that $\hat a(x) = (\hat a_0(x), \dots, \hat a_p(x))^T$ satisfies the equation $$(X_x^TW_xX_x)\hat a(x) = X_x^T W_xY\eqno{(4.4)}$$
We use Lemma 1 for implementation of the QGJE code to solve the equation (4.4) on QNN, see also \cite{dndiep1} and hence to implement the KNR-QNN. In particular we have $\hat r_n(x) = \hat a_0(x)$

For simplify the notation we consider the case of polynomial regression with one unknown variable. The several-variable case is similar by using the multi-index notations.

The following result was shown \cite{wasserman}:
\textit{The estimator for a local polynomial regression is
$\hat r_n(x) = \sum_{i=1}^n \ell_i(x)y_i$, where $\ell(x) = (\ell_1(x),\dots,\ell_n(x))^T$ can be find out from the equation
$$X_x^TW_xX_x\hat a(x) = X_x^TW_xY$$ with expectation 
$\mathbb E(\hat r_n(x)) = \sum_{i=1}^n \ell_i(x)r(x_i)$, and variance $\mathbb V(\hat r_n(x) = \sigma^2\sum_{i=1}^n \ell_i(x)^2 = \sigma^2||\ell(x)||.$
}

Let us use the confidence intervals $I(x)$ as the training rule.

\section{Conclusion}
We implemented the quantum neural networks: the nonparametric least square quantum neural network NLS-QNN, the polynomial nonparametric regression quantum neural network PNR-QNN by using the Grover's Search quantum neural network GS-QNN and Gauss-Jordan Elimination quantum neural network GJE-QNN.  The training rules are provided with the corresponding high probability intervals from Nonpaprametric Statistics.

\section{Acknowledgments}
The authors express their sincerle thanks to Professor H. Geurdes, Professor S. Heidari and Professor G. Resconi for valuable comments.

\vskip 0.25cm
\textsc{Note.} On behalf of all the authors the corresponding author declaires that there is no conflict of interest.


\begin{thebibliography}{xxx} 



\bibitem{hinton}{\sc G. Hinton}, {\it A Practical Guide to Training Restricted Boltzmann Machines}, 
Department of Computer Science, University of Toronto. 

\bibitem{wiebeetal}{\sc N. Wiebe, A. Kapoor, K. Swore}, {Quantum deep learning}, arXiv:1412.3489v2[quant-ph], 2015.


\bibitem{uvarovetal}{\sc A. V. Uvarov, A.S. Kardeshin, J.D. Biamonte}, {\it Machine learning phase transition with a quantum processor}, arXiv:1906.10155v1[quant-ph], 2019.

\bibitem{schuldetal}{\sc M. Schuld, I. Sinayskiy, Petruccione}, {\it An introduction to quantum machine learning}, arXiv:1409.3097v1[quant-ph]2014.

\bibitem{dunjkobriegel}{\sc }{\sc V. Dunjko, H. J. Briegel}, {\it Machine learning \& artificial intelligence in the quantum domain}, arXiv:17090277v1[quant-ph], 2017.

\bibitem{wiebewossnig}{\sc N. Wiebe, L. Wossnig}, {\it Generative training of quantum Boltzmann machines with hidden units}, arXiv:190509902v1[quant-ph], 2019.

\bibitem{ezhovventura}{\sc A. A. Ezhov,  D. Ventura}, {\it Quantum neural networks}, in Future \textit{Directions for Intelligent Systems and Information Science},  N. Kasabov (ed.), Physica-Verlag, pp. 213-235, 2000.





\bibitem{wasserman}{\sc L. Wasserman}, {\it  All of Nonparametric Statistics}, Springer
Berlin Heidelberg New York Barcelona Hong Kong London Milan Paris, pp.278, 2005.
Tokyo

\bibitem{dndiep1}{\sc D. N. Diep, D. H. Giang, N. V. Minh},  {\it Quantum
Gauss-Jordan elimination and simulation ofnaccounting principles
 on quantum computers}, Inter. J. of Theor. Physics, 56(2017), No 6, 1948-1960.

\bibitem{dndiep2}{\sc K. Nagata, S. K. Patro, H. Geurdes, S. Heidari, D. N. Diep, T. Nakamura}, {\it Various New Forms of the Bernstein-Vazirani Algorithm Beyond Qubit Systems},  Asian J. Math. \& Phys., 3, No 1(2019) 1-12.

\bibitem{dndiep3}{\sc D. N. Diep}, {\it Some quantum neural networks}, Intl. J. Theor. Phys. (to appear).

\bibitem{dndiep4}{\sc D. N. Diep}, {\it Statistical tests and confidential intervals as thresholds for quantum neural networks}, Quantum Machine Intelligence (to appear).


\bibitem{guptazia}{\sc S. Gupta, R.K. P. Zia}, {\it  Quantum Neural Network}, Journal of computer and system sciences, 63(2001), 355-383.

\end{thebibliography}
\end{document}